\documentclass[3p,twocolumn]{elsarticle}

\PassOptionsToPackage{numbers, compress}{natbib}
\usepackage[utf8]{inputenc} 
\usepackage[T1]{fontenc}    
\usepackage{url}            
\usepackage{booktabs}       
\usepackage{amsfonts}       
\usepackage{nicefrac}       
\usepackage{microtype}      
\usepackage{xcolor}         
\usepackage{yhmath}

\usepackage{bm}             
\usepackage{amsmath}
\usepackage{amssymb}
\usepackage{cprotect}
\usepackage{dblfloatfix}

\newcommand{\blank}{\mathrel{\;\cdot\;}}

\DeclareMathOperator*{\argmin}{argmin}

\newcommand{\lm}{\textcolor{black}}

\newcommand\corestr[2]{{
  \left.\kern-\nulldelimiterspace 
  #1
  \vphantom{\big|}
  \right|^{#2}
}}

\begin{document}

\title{Super-resolving sparse observations in partial differential equations:\\ A physics-constrained convolutional neural network approach}

\author[1]{Daniel Kelshaw}
\ead{djk21@ic.ac.uk}

\author[1,2]{Luca Magri\corref{cor1}}
\ead{l.magri@imperial.ac.uk}

\cortext[cor1]{Corresponding Author}

\affiliation[1]{
    organization={Department of Aeronautics, Imperial College London},
    addressline={Exhibition Road},
    city={London},
    postcode={SW7 2AZ},
    country={United Kingdom}
}

\affiliation[2]{
    organization={Alan Turing Institute},
    addressline={British Library, 96 Euston Rd.},
    city={London},
    postcode={NW1 2DB},
    country={United Kingdom}
}

\begin{abstract}
    We propose the physics-constrained convolutional neural network (PC-CNN) to
    infer the high-resolution solution from sparse observations of spatiotemporal and nonlinear partial differential equations. Results are shown for a chaotic and turbulent fluid motion, whose solution is high-dimensional, and has fine spatiotemporal scales.
    We show that, by constraining prior physical knowledge in the CNN, we can infer the unresolved physical dynamics without using the high-resolution dataset in the training. 
    This opens opportunities for super-resolution of experimental data and low-resolution simulations. 
\end{abstract}

\maketitle

\section{Introduction}
Observations of turbulent flows and physical systems are limited in many cases, with only sparse or partial measurements being accessible. Access
to limited information obscures the underlying dynamics and provides a challenge for system identification \citep[e.g.,][]{Brunton2016}. Super-resolution
methods offer the means for high-resolution state reconstruction from limited observations, which is a key objective for experimentalists and computational
scientists alike.
Current methods for super-resolution or image reconstruction primarily make use of convolutional neural networks due to their ability to exploit
spatial correlations. In the classic, data-driven approach there is a requirement for access to pairs of low- and high-resolution samples, which are needed to produce 
a parametric mapping for the super-resolution task~\citep[e.g.,][]{Dong2014, Shi2016, Yang2019, Liu2020}. Considering observations of physical systems, the
problems encountered in the absence of ground-truth high-resolution observations can be mitigated by employing a physics-constrained approach by imposing
prior knowledge of the governing equations~\citep[e.g.,][]{Lagaris1998,Raissi2019,doan2021short}.

Physics-informed neural networks (PINNs)~\citep{Raissi2019} have provided a tool for physically-motivated problems, which exploits 
automatic-differentiation to constrain the governing equations. 
Although the use of PINNs for super-resolution shows promising results for simple systems~\citep{Eivazi2022}, they remain challenging to train, and are not designed to exploit spatial correlations~\citep{krishnapriyan2021characterizing, grossmann2023can}. On the other hand, convolutional neural networks are designed to exploit spatial correlations, but they cannot naturally leverage automatic differentiation to evaluate the physical loss, as PINNs do, because they provide a mapping between states, rather than a mapping from the spatiotemporal coordinates to states as in PINNs. As such, 
the design of physics-informed convolutional neural networks is more challenging, and rely on finite-different approximations or differentiable 
solvers~\citep{kelshaw2023uncovering, Gao2021}. 
For example, the authors of \citep{Gao2021} show results for a steady flow field (with no temporal dynamics), which produces a mapping for stationary solutions of the Navier-Stokes equations. In this work, we design a framework to tackle spatiotemporal partial differential equations.
\begin{figure*}[b!]
    \centering
    \includegraphics[width=\linewidth]{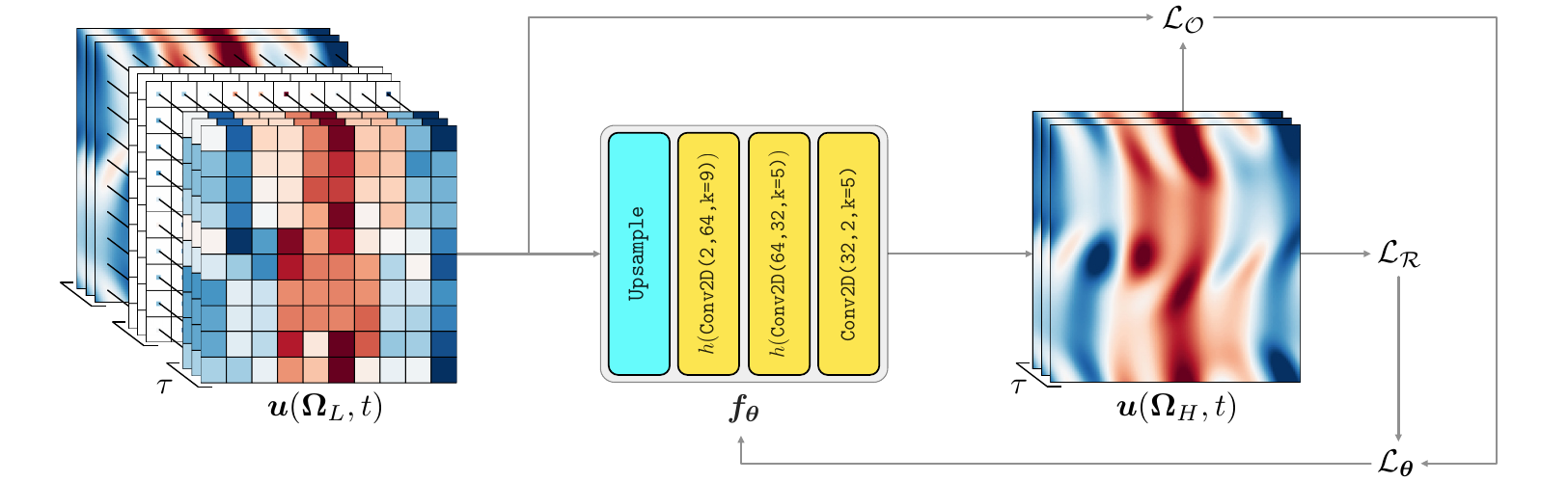}
    \cprotect
    \caption{Super-resolution task. The model ${\bm {f_\theta}}$ is responsible for mapping the low-resolution field ${\bm u}({\bm \Omega_L}, t)$
        to the high-resolution field ${\bm u}({\bm \Omega_H}, t)$. The upsampling layers (middle white layers in the leftmost set of data) perform bi-cubic upsampling to obtain the
        correct spatial dimensions. Convolutional layers are parameterised as \verb|Conv2D(c_in, c_out, k)|, where \verb|c_in, c_out| denote the
        number of input and output channels, respectively; and \verb|k| is the spatial extent of the filter. The terms $\mathcal{L}_{\mathcal{O}}, 
        \mathcal{L}_{\mathcal{R}}$ denote the observation-based loss and residual loss, respectively, the combination of which forms the objective loss
        $\mathcal{L}_{\bm \theta}$, which is used to update the network's parameters, ${\bm \theta}$. We provide an in-depth explanation of these
        losses in §\ref{sec:loss_terms}.
    }
    \label{fig:overview}
\end{figure*}
For this, we propose a super-resolution method that does not require the full set of high-resolution observations. To accomplish this, we design and propose
the physics-constrained convolutional neural network (PC-CNN) with a time windowing scheme. We apply this to a two-dimensional chaotic and turbulent fluid motion.
The paper is structured as follows.
In §\ref{sec:methodology} we introduce the super-resolution task in the context of a mapping from low- to high- resolution. The choice of architecture 
used to represent the mapping is discussed in §\ref{sec:cnns} before providing an overview of the methodology §\ref{sec:loss_terms}. We introduce
the two-dimensional, turbulent flow in §\ref{sec:kolmogorov} and provide details on the pseudospectral method used for discretisation. Finally, results 
are demonstrated in §\ref{sec:results}. Conclusions~\S\ref{sec:conclusion} end the paper.
 
\section{Super-resolution task} \label{sec:methodology}
We consider partial differential equations (PDEs) of the form
\begin{equation} \label{eqn:dynamical_system}
    \mathcal{R}({\bm {\tilde u}}; \lambda) \equiv \partial_t {\bm {\tilde u}} - \mathcal{N}({\bm {\tilde u}}; \lambda),
\end{equation}
\lm{
where ${\bm x} \in \Omega \subset \mathbb{R}^n$ denotes the spatial location; 
$t \in [0, T] \subset \mathbb{R}_{\geq 0}$ is the time, where $n$ is the space dimension; 
${\bm {\tilde u}}: \Omega \times [0, T] \rightarrow \mathbb{R}^{m}$ is the state, where where $m$ is the number of components of the vector function ${\bm {\tilde u}}$}; 
$\lambda$ are the physical parameters of the system; $\mathcal{N}$ is a 
sufficiently smooth differential operator, which represents the nonlinear part of the system of $m$ partial differential equations; 
$\mathcal{R}$ is the residual; 
and $\partial_t$ is the partial derivative with respect to time.
A solution ${\bm u}$ of the PDE~\eqref{eqn:dynamical_system} is the function that makes the residual vanish, i.e., $\mathcal{R}({\bm u}; \lambda) = 0$.

Given sparse observations, ${\bm u}({\bm \Omega_L}, t)$, we aim to reconstruct the underlying solution to the partial differential equation 
on a high-resolution grid, ${\bm u}({\bm \Omega_H}, t)$, where the domain $\Omega$ is discretised on low- and high-resolution \lm{uniform} grids, \lm{
${\bm \Omega_L} \subset \mathbb{R}^{N^n}$} and \lm{${\bm \Omega_H} \subset \mathbb{R}^{M^n}$}, respectively, such that 
${\bm \Omega_L} \cap {\bm \Omega_H} = {\bm \Omega_L}$; $M = \kappa N$; and $\kappa \in \mathbb{N}^{+}$ is the up-sampling factor.  Our objective
is to find a parameterised mapping ${\bm {f_\theta}}$ such that
\begin{equation}
    {\bm {f_\theta}}: {\bm u}({\bm \Omega_L}, t) \rightarrow {\bm u}({\bm \Omega_H}, t). 
\end{equation}
We consider the solution, ${\bm u}$, to be discretised with $N_t$ times steps, $t_i$, which are contained in the set $\mathcal{T} = \{ t_i \in [0, T] \}_{i=0}^{N_t}$. We approximate
the mapping ${\bm {f_\theta}}$ by a convolutional neural network parameterised by ${\bm \theta}$.

\section{Convolutional neural networks} \label{sec:cnns}
Parameterising ${\bm {f_\theta}}$ as a convolutional neural network is a design choice, which allows us to exploit two key 
properties. First, shift invariance, which is consequence of kernel-based operations acting on structured grids~\citep{Gu2018}.
Second, partial differential equations are defined by local operators, which is a property that we 
wish the machine to induce naturally. Therefore, we choose convolutional neural networks, which employ an architectural paradigm that leverages spatial correlations~\citep{lecun1995convolutional}, through the composition of functions 
\begin{equation}
    {\bm {f_\theta}} = {\bm f}^{Q}_{\bm \theta_Q} \circ \cdots \circ h({\bm f}^{1}_{\bm \theta_1}) \circ h({\bm f}^{0}_{\bm \theta_0}), 
\end{equation}
where ${\bm f}^i_{{\bm \theta}_i}$ denote discrete convolutional layers;  $h$ is an element-wise nonlinear activation function, which 
increases the expressive power of the network~\citep[e.g.,][]{magri2023notes} and yields a universal function approximator~\citep{Hornik1989}; and $Q$ is the number of layers. The convolutional layers are
responsible for the discrete operation $({\bm x}, \bm{w}, \bm{b}) \mapsto {\bm w} \ast {\bm x} + {\bm b}$, where 
${\bm \theta} = ({\bm w}, {\bm b})$. As the kernel operates locally around each pixel, information is leveraged  from the surrounding grid 
cells. This makes convolutional layers an excellent choice for learning and exploiting spatial correlations, as are often important in the
solutions to partial differential equations~\citep[e.g.,][]{Murata2020, Gao2021}.

We propose a physics-constrained convolutional neural network (PC-CNN), which consists of three successive convolutional layers that are prepended by an 
upsampling operation tasked with increasing the spatial resolution of the input. 
Knowledge of the boundary conditions can be imposed through
the use of padding; for instance periodic padding for periodic boundary conditions. Figure~\ref{fig:overview} provides an overview of the 
super-resolution task with information about the architecture employed.
%

\section{Methodology} \label{sec:loss_terms}
Having defined the mapping ${\bm {f_\theta}}$ as a convolutional neural network parameterised by ${\bm \theta}$, we formalise an optimisation problem
to minimise the cost function $\mathcal{L}_{\bm \theta}$
\begin{equation} \label{eq:f3iefd4230432-dd}
    {\bm \theta}^\ast = \argmin_{\bm \theta} \mathcal{L}_{\bm \theta} \quad \text{where} \quad
    \mathcal{L}_{\bm \theta} = \mathcal{L}_{\mathcal{O}} + \alpha \mathcal{L}_{\mathcal{R}},
\end{equation}
where $\alpha$ is a non-negative empirical regularisation factor, which determines the relative importance of the corresponding loss terms. Given 
low-resolution observations ${\bm u}({\bm \Omega_L}, t)$ at arbitrary times $t \in \mathcal{T}$, we define each of the loss terms as 
\begin{equation} \label{eqn:lo_lr}
\begin{aligned}[c]
    \mathcal{L}_{\mathcal{O}} &= \frac{1}{N_t} \sum_{t \in \mathcal{T}} \lVert 
        \corestr{{\bm {f_\theta}}({\bm u}({\bm \Omega_L}, t))}{{\bm \Omega_L}} - {\bm u}({\bm \Omega_L}, t)
    \rVert_{{\bm \Omega_L}}^{2}, \\
    \mathcal{L}_{\mathcal{R}} &= \frac{1}{N_t} \sum_{t \in \mathcal{T}} \lVert 
       \mathcal{R}({\bm {f_\theta}}({\bm u}({\bm \Omega_L}, t)); \lambda)
    \rVert_{{\bm \Omega_H}}^{2},
\end{aligned}
\end{equation}
where $\corestr{{\bm {f_\theta}}(\blank)}{{\bm \Omega_L}}$ denotes the corestriction of ${\bm \Omega_H}$ on ${\bm \Omega_L}$, and 
$\lVert \blank \rVert_{\bm \Omega}$ represents the $\ell^2$-norm over the domain ${\bm \Omega}$. In order to find an optimal set of parameters
${\bm \theta}^\ast$, the loss is designed to regularise predictions that do not conform to the desired output. Given sensor observations
${\bm u}({\bm \Omega_L}, t)$ in the low-resolution input, the observation-based loss, $\mathcal{L}_{\mathcal{O}}$, is defined to minimise the distance
between known observations,  ${\bm u}({\bm \Omega_L}, t)$, and their corresponding predictions on the high-resolution grid,  
${\bm {f_\theta}}({\bm u}({\bm \Omega_L}, t))$. 

We impose the prior knowledge that we have on the dynamical system by defining the residual loss, $\mathcal{L}_{\mathcal{R}}$, which penalises the parameters that yield
predictions that violate the governing equations~\eqref{eqn:dynamical_system}\footnote{For example, in turbulence, the fluid mass and momentum must be in balance with mass sources, and forces, respectively. Thus, \eqref{eqn:dynamical_system}  represent conservation laws.}. 
Mathematically, this means that weIn absence of observations (i.e., data), the residual loss $\mathcal{L}_{\mathcal{R}}$ 
alone does not provide a unique solution. By augmenting the residual loss with the data loss, $\mathcal{L}_{\mathcal{O}}$ (see Eq.~\eqref{eq:f3iefd4230432-dd}), 
we ensure that network realisations conform to the observations (data) whilst fulfilling the governing equations, e.g., conservation laws.
Crucially, as consequence of the proposed training objective, we do not need the high-resolution field as labelled dataset, which is required 
with conventional super-resolution methods. 

\subsection{Time-windowing of the residual loss}
Computing the residual of a partial differential equation is a temporal task, as shown in Eq.~\eqref{eqn:dynamical_system}.
We employ a time-windowing approach to allow the network ${\bm {f_\theta}}$ to learn the sequentiality of the data.  This provides
a means to compute the time-derivative $\partial_t {\bm u}$ required for the residual loss $\mathcal{L}_{\mathcal{R}}$. The network
takes time-windowed samples as inputs, each sample consisting of $\tau$ sequential time-steps. The time-derivative $\partial_t {\bm u}$ 
is computed by applying a forward Euler approximation to the  
loss
\begin{equation}
\begin{split}
    \mathcal{L}_{\mathcal{R}} = \frac{1}{\tau N_t} \sum_{t \in \mathcal{T}} \sum_{n=0}^{\tau}
    \big\lVert 
            \partial_{t} {\bm {f_\theta}}({\bm u}({\bm \Omega_L}, t + n \Delta t))
        \\- 
            \mathcal{N}({\bm {f_\theta}}({\bm u}({\bm \Omega_L}, t + n \Delta t)); \lambda)
    \big\rVert_{{\bm \Omega_H}}^{2}.
    \end{split}
\end{equation}
Using this approach, we are able to obtain the residual for the predictions in a temporally local sense; computing derivatives across 
discrete time windows rather than the entire simulation domain. The network ${\bm {f_\theta}}$ is augmented to operate on these time windows,
which vectorises the operations over the time-window.
For a fixed quantity of training data, the choice of $\tau$ introduces a trade-off between the number of input samples $N_t$,
and the size of each time-window $\tau$. We take a value $\tau = 2$, which is the minimum window size  for computing the 
residual loss $\mathcal{L}_{\mathcal{R}}$. We find that this is sufficient for training the network whilst simultaneously maximising 
the number of independent samples used for training. To avoid duplication of the data in the training set, we ensure that all samples are at 
least $\tau$ time-steps apart so that independent time-windows are guaranteed to not contain any overlaps.

\section{Chaotic and turbulent dataset} \label{sec:kolmogorov}
As nonlinear partial differential equations, we consider the Navier-Stokes equations, which are the expressions of the conservation of mass and momentum of fluid motion, respectively 
\begin{equation}
\begin{split}
    \nabla \cdot {\bm u} &= 0, \\
    \partial_t {\bm u} + \left( {\bm u} \cdot \bm{\nabla} \right) {\bm u} &= - \nabla p + \nu \Delta {\bm u} + \bm{g},
\end{split}
\end{equation}
where $p, \nu$ denote the scalar pressure field and kinematic viscosity, respectively. 
\lm{The flow velocity,  ${\bm u}\in \mathbb{R}^{m=2}$ evolves on the domain 
$\Omega \in [0, 2 \pi) \subset \mathbb{R}^{n=2}$} with periodic boundary conditions applied on $\partial \Omega$, and a stationary, spatially-varying
sinusoidal forcing term ${\bm g}({\bm x})$. In this paper, we take $\nu = \nicefrac{1}{42}$ to ensure chaotic and turbulent dynamics, and employ
the forcing term ${\bm g}({\bm x}) = \left[ \sin{(4\bm{x_2}), 0} \right]^\top$~\citep{Fylladitakis2018}, where $\bm{x_2}$ is the transverse coordinate.
This flow, which is also known as the Kolmogorov flow~\citep{Fylladitakis2018}, provides a nonlinear and multi-scale dateset, which allows us to evaluate the 
quality of predictions across the turbulent spectrum.  

\subsection{Differentiable pseudospectral discretisation} \label{sec:spectral}
To produce a solution for the Kolmogorov flow, we utilise a differentiable pseudospectral spatial discretisation 
$\bm{\hat{\bm u}}_k = \hat{\bm u}({\bm k}, t)$ where $\hat{\bm u} = \mathcal{F} \circ {\bm u}$; $\mathcal{F}$ is the Fourier transform; and 
\lm{${\bm k} \in \hat{\bm \Omega}_k \subset \mathbb{C}^{K^n}$} is the spectral discretisation of the spatial domain $\Omega \in [0, 2\pi)$~\citep{kolsol2022}.
Operating in the Fourier domain eliminates the continuity term \citep{Canuto1988}. The equations for the spectral representation of the
Kolmogorov flow are
\begin{equation}
    \mathcal{R}(\bm{\hat{\bm u}}_k; \lambda) = \left( \tfrac{d}{dt} + \nu \lvert {\bm k} \rvert^2 \right) \bm{\hat{\bm u}}_k - {\bm {\hat f}}_k + {\bm k} \frac{{\bm k} \cdot {\bm {\hat f}}_k}{\lvert {\bm k} \rvert^2} - {\hat {\bm g}}_k,
\end{equation}
with ${\bm {\hat f}}_k = - \left( \widehat{{\bm u} \cdot \nabla {\bm u}} \right)_{k}$,
where nonlinear terms are handled pseudospectrally, employing the $\nicefrac{2}{3}$ de-aliasing rule to avoid unphysical
culmination of energy at the high frequencies~\cite{Canuto1988}.
A solution is produced by time-integration of the dynamical system with the explicit forward-Euler scheme, choosing a time-step $\Delta t$ that satisfies
the Courant-Friedrichs-Lewy (CFL) condition. Initial conditions are generated by producing a random field scaled by the wavenumber,
which retains the spatial structures of varying lengthscale in the physical domain~\citep{ruan1998}. The initial transient, which is approximately $T_t=180s$, is removed from the dataset to ensure that the results are
statistically stationary. (The  transient time is case dependant.)
\lm{For a spatial resolution ${\bm \Omega_H} \in \mathbb{R}^{70 \times 70}$, we use a spectral discretisation 
$\hat{\bm \Omega}_k \in \mathbb{C}^{35 \times 35}$} to avoid aliasing and resolve the smallest lengthscales possible.

\begin{figure*}[b!]
    \centering
    \includegraphics[width=\linewidth]{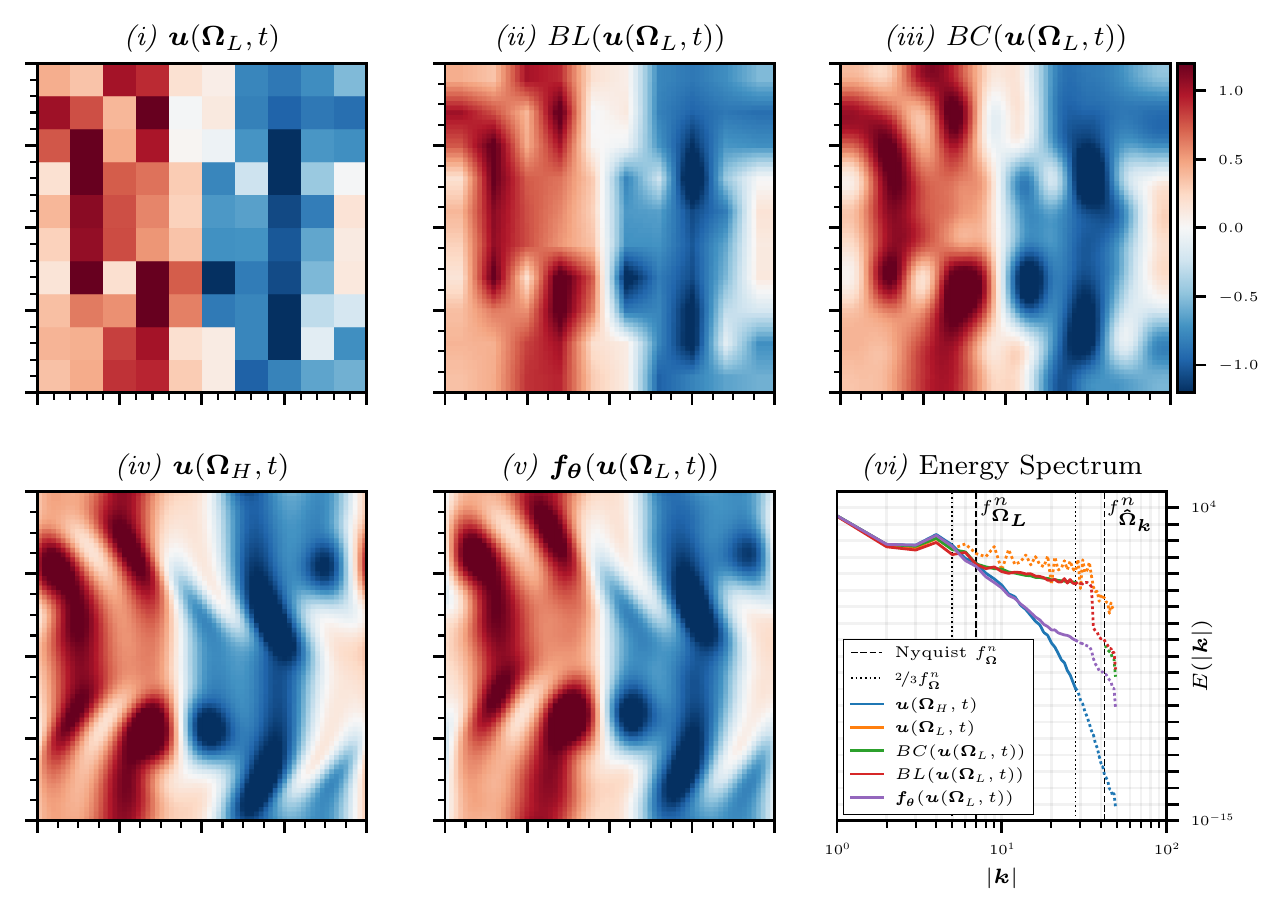}
    \caption{Super-resolution results. Physics-constrained convolutional neural network (PC-CNN) compared with interpolation methods. 
        Panel \textit{(i)}~shows the low-resolution input, ${\bm u}({\bm \Omega_L}, t)$;
        \textit{(ii)}~bi-linear interpolation, $\textit{BL}({\bm u}({\bm \Omega_L}, t))$;
        \textit{(iii)}~bi-cubic interpolation, $\textit{BC}({\bm u}({\bm \Omega_L}, t))$; 
        \textit{(iv)}~true high-resolution field, ${\bm u}({\bm \Omega_H}, t)$; 
        \textit{(v)}~model prediction of the high-resolution field, ${\bm {f_\theta}}({\bm u}({\bm \Omega_L}, t))$; 
        and \textit{(vi)}~energy spectra for each of the predictions.
    }
    \label{fig:interpolation}
\end{figure*}

\subsection{Residual loss in the Fourier domain}
The pseudospectral discretisation provides an efficient means to compute the differential operator $\mathcal{N}$, which allows  
us to evaluate the residual loss $\mathcal{L}_{\mathcal{R}}$ in the Fourier domain
\begin{equation}
\begin{split}
    \mathcal{L}_{\mathcal{R}} = \frac{1}{\tau N_t} \sum_{t \in \mathcal{T}} \sum_{n=0}^{\tau} \lVert
        \partial_t \hat{\bm {f_\theta}}({\bm u}({\bm \Omega_L}, t + n \Delta t)) \\- \hat{\mathcal{N}}(\hat{\bm {f_\theta}}({\bm u}({\bm \Omega_L}, t + n \Delta t)))
    \rVert_{\bm{\hat{\Omega}_{k}}}^{2},
\end{split}
\end{equation}
where $\hat{\bm {f_\theta}} = \mathcal{F} \circ {\bm {f_\theta}}$, and $\hat{\mathcal{N}}$ denotes the Fourier-transformed
differential operator. The pseudospectral discretisation is fully differentiable, which allows us to numerically 
compute gradients with respect to the parameters of the network ${\bm {f_\theta}}$. Computing the loss $\mathcal{L}_{\mathcal{R}}$ in 
the Fourier domain provides two advantages: \textit{(i)} periodic boundary conditions are naturally enforced, which enforces the  prior knowledge in the loss calculations; and \textit{(ii)} gradient calculations yield 
spectral accuracy. In contrast, a conventional finite differences approach requires a computational stencil, the spatial extent of 
which places an error bound on the gradient computation. This error bound is a function of the spatial resolution of the field. 

\section{Results} \label{sec:results} 
First, we discuss the generation of the low-resolution data for the super-resolution task. Next, we show the ability of the PC-CNN
to infer the high-resolution solution of the partial differential equation on points that are not present in the training set. 

\subsection{Obtaining the low-resolution data}
A high-resolution solution of the partial differential equation is generated on the grid ${\bm \Omega_H}$ prior to extracting
a low-resolution grid ${\bm \Omega_L}$ with the downsampling factor of $\kappa = \nicefrac{N}{M}$.
%
%
Both the solver and residual loss are discretised with $K = \nicefrac{N}{2}$ wavenumbers in the Fourier domain, which  complies 
with the Nyquist-Shannon sampling criterion. The downsampling by $\kappa$ is performed by extracting the value of the solution 
at spatial locations ${\bm \Omega_L} \cap {\bm \Omega_H}$, i.e.,  a low-resolution representation of the high-resolution  solution 
\begin{equation}
    {\bm u}({\bm \Omega_L}, t) \triangleq \corestr{{\bm u}({\bm \Omega_H}, t)}{{\bm \Omega_L}}.
\end{equation}
(In contrast, the use of a pooling method for downsampling would distort values in the low-resolution representation, which effectively 
modifies the high-resolution solution.) 

\subsection{Comparison with standard upsampling}
We showcase the results for a downsampling factor $\kappa = 7$. Results are compared with interpolating upsampling methods, i.e., bi-linear, and bi-cubic interpolation 
to demonstrate the ability of the method. We provide a notion of quantitative accuracy by computing the relative $\ell^2$-error between the 
true solution, ${\bm u}({\bm \Omega_H}, t)$, and the corresponding network realisation, ${\bm {f_\theta}}({\bm u}({\bm \Omega_L}, t))$
\begin{equation} \label{eqn:error}
    e = 
    \sqrt{
        \frac{
            \sum_{t \in \mathcal{T}} \lVert {\bm u}(\bm{\Omega_H}, t) - {\bm {f_\theta}}({\bm u}(\bm{\Omega_L}, t)) \rVert_{\bm{\Omega_H}}^{2}
        }{
            \sum_{t \in \mathcal{T}} \lVert {\bm u}(\bm{\Omega_H}, t) \rVert_{\bm{\Omega_H}}^{2}
        }
    }.
\end{equation}
Upon discarding the transient, a solution ${\bm u}({\bm \Omega_H}, t) \in \mathbb{R}^{70 \times 70}$ is generated by
time-integration over $12 \times 10^3$ time-steps, with $\Delta t = 1 \times 10^{-3}$.  We extract the low-resolution 
solution ${\bm u}({\bm \Omega_L}, t) \in \mathbb{R}^{10 \times 10}$ as a candidate for the super-resolution task. We extract $2048$ samples
at random from the time-domain of the solution, each sample consisting of $\tau = 2$ consecutive time-steps. The  \textit{adam} optimiser~\citep{kingma2015} is 
employed for training with a learning rate of $3 \times 10^{-4}$. We take $\alpha = 10^{3}$ as the regularisation factor for the loss 
$\mathcal{L}_{\bm \theta}$ and train for a total of $10^{3}$ epochs, which is empirically determined to provide sufficient convergence.

Figure \ref{fig:interpolation} shows a snapshot of results for the streamwise component of the velocity field, comparing network realisations
${\bm {f_\theta}}({\bm u}({\bm \Omega_L}, t))$ with the interpolated alternatives. Bi-linear and bi-cubic interpolation are denoted by 
$BL({\bm u}({\bm \Omega_L}, t)), BC({\bm u}({\bm \Omega_L}, t))$ respectively. We observe that network realisations of the high-resolution solution
yield qualitatively more accurate results as compared with the interpolation. Artefacts indicative of the interpolation scheme 
used are present in both of the interpolated fields, whereas the network realisation captures the structures present in the high-resolution 
field correctly. Across the entire solution domain the model ${\bm {f_\theta}}$ achieves a relative $\ell^2$-error 
of $e=3.449 \times 10^{-2}$ compared with $e=2.091 \times 10^{-1}$ for bi-linear interpolation and $e=1.717 \times 10^{-1}$ for bi-cubic interpolation.

Although the relative $\ell^2$-error provides a notion of predictive accuracy, it is crucial to assess the 
physical characteristics of the super-resolved field~\citep{pope2000turbulent}. The energy spectrum, which is characteristic of turbulent flows, 
represents a multi-scale phenomenon where energy content decreases with the wavenumber. From the energy spectrum of the network's 
prediction, ${\bm {f_\theta}}({\bm u}(\bm{\Omega}_L, t))$, we gain physical insight into the multi-scale nature of the solution. Results 
in Figure \ref{fig:interpolation} show that the energy content of the low-resolution field diverges from that of the high-resolution
field, which is a consequence of spectral aliasing. Network realisations ${\bm {f_\theta}}({\bm u}({\bm \Omega_L}, t))$ are capable of capturing 
finer scales of turbulence compared to both interpolation approaches, prior to diverging from the true spectrum as $\lvert {\bm k} \rvert = 18$. 
The residual loss, $\mathcal{L}_{\mathcal{R}}$, enables the network to act beyond simple interpolation. The network is capable of 
de-aliasing, thereby inferring unresolved physics. Parametric studies show similar results across a range of super-resolution factors
$\kappa$ (result not shown).

\section{Conclusions} \label{sec:conclusion}
In this paper, we introduce a method for physics-constrained super-resolution of observations in partial differential equations without
access to the full high-resolution samples. 
First, we define the super-resolution task and introduce the physics-constrained convolutional
neural network (PC-CNN), which provides the framework to compute physical residuals for spatiotemporal systems. 
Second, we formulate an optimisation
problem by leveraging knowledge of the partial differential equations and low-resolution observations to regularise the predictions from the 
 network. 
Third, we showcase the PC-CNN on a turbulent flow, which is a spatiotemporally chaotic solution of the nonlinear partial differential equations of fluid motion (Navier-Stokes). 
Finally, we demonstrate that the proposed PC-CNN 
provides more accurate physical results, both qualitatively and quantitatively, as compared to interpolating upsampling methods.
This work opens opportunities for the accurate reconstruction of solutions of partial differential equations from sparse observations, as is prevalent in experimental settings, without the full set of high-resolution images.

\section{Acknowledgements}
D. Kelshaw. and L. Magri. acknowledge support from the UK Engineering and Physical Sciences Research Council. 
L. Magri gratefully acknowledges financial support from the ERC Starting Grant PhyCo 949388.

\bibliography{references}

\end{document}